\documentclass[review]{elsarticle}
\usepackage{graphicx}
\usepackage{rotating}
\usepackage{tabularx}
\usepackage{float}
\usepackage{svg}
\usepackage{amsmath}
\usepackage{setspace}
\graphicspath{./imgs/}

\usepackage{hyperref}

\journal{Elsevier}

\biboptions{authoryear}

\begin{document}

\begin{frontmatter}

\title{Exploring Visual Complaints through a test battery in Acquired Brain Injury Patients: A Detailed Analysis of the DiaNAH Dataset}



\author{Gonçalo Hora de Carvalho}
\address{}
\ead[url]{goncalo.horacarvalho@gmail.com}




\begin{abstract}
This study investigated visual impairment complaints in a sample of 948 Acquired Brain Injury (ABI) patients using the DiaNAH dataset, emphasizing advanced machine learning techniques for managing missing data. Patients completed a CVS questionnaire capturing eight types of visual symptoms, including blurred vision and altered contrast perception. Due to incomplete data, 181 patients were excluded, resulting in an analytical subset of 767 individuals. To address the challenge of missing data, an automated machine learning (AutoML) approach was employed for data imputation, preserving the distributional characteristics of the original dataset. Patients were grouped according to singular and combined complaint clusters derived from the 40,320 potential combinations identified through the CVS questionnaire. A linear correlation analysis revealed minimal to no direct relationship between patient-reported visual complaints and standard visual perceptual function tests. This study represents an initial systematic attempt to understand the complex relationship between subjective visual complaints and objective visual perceptual assessments in ABI patients. Given the limitations of sample size and variability, further studies with larger populations are recommended to robustly explore these complaint clusters and their implications for visual perception following brain injury.

\end{abstract}


\end{frontmatter}

\section*{Introduction}
The human brain, in its incredible complexity, orchestrates the perception, understanding, and interaction with the world around us. The capacity of this organ to process vast quantities of information is facilitated, in no small part, by our sense of vision. However, injury to the brain can have a significant impact on visual perception and related functions and previous studies have demonstrated that visual symptoms are prevalent among Acquired Brain Injuries (ABIs) patients and can persist long after the initial injury (Wang et al., 2021). This study seeks to further explore the utility of the DiaNAH dataset - a visual test battery on ABI patients.


One of the challenges of studies involving human participants is the construction of consistent and comprehensive datasets. Researchers also often face the issue of missing data, which can occur for many reasons, from patients not completing certain sections of a survey to technical glitches in data collection apparatuses (Smith \& Jones, "Challenges in Medical Data Collection").

The field of ABIs has seen considerable growth in the last decade with the advent of datasets such as the Traumatic Brain Injury Model Systems (TBIMS) National Database (Miller et al., "The TBIMS National Database: Advancements and Future Directions"). However, the DiaNAH dataset, with its focus on visual complaints post-ABI, occupies a unique niche. 

Given its size, of the main challenges in utilizing the DiaNAH dataset is dealing with missing data. Missing data can occur for various reasons, including incomplete survey responses and technical issues during data collection. This issue can significantly impact the reliability of the analysis if not appropriately addressed. Traditional methods for handling missing data, such as list-wise deletion or mean imputation, can lead to biased results and reduced statistical power (Rubin, 2004). Advanced machine learning techniques for data imputation may provide a more reliable alternative, preserving the dataset's integrity and allowing for more accurate analysis (Van Buuren \& Groothuis-Oudshoorn, 2011).

These machine learning algorithms can predict and fill in missing values based on the available data, thereby maintaining the dataset's overall structure and variability. This approach is particularly useful in medical research, where incomplete data is a common challenge (Sterne et al., 2009).

The analysis of the DiaNAH dataset involved then examining the relationships between visual complaints and various visual perceptual function tests. This study aimed to identify potential clusters of complaints and explore their correlations with the test results. Our motivation to do this was to understand these relationships so as to deduce underlying mechanisms of visual impairments in ABI patients and inform the development of targeted interventions.

Previous research has highlighted the complexity of visual complaints in ABI patients, suggesting that these complaints may not always correlate directly with measurable visual deficits (Geldmacher et al., 2020). For instance, a study by Miller et al. (2019) found that while many ABI patients report significant visual complaints, these are not always reflected in standard visual perceptual tests. This discrepancy underscores the need for a more nuanced approach to analyzing visual complaints, considering both subjective reports and objective test results.

The DiaNAH dataset, together with a CVS questionnaire and extensive battery of visual tests, provides a unique opportunity to delve deeper into these issues. By employing data analysis techniques and machine learning algorithms, this study seeks to uncover patterns and correlations that might not be apparent through traditional methods. The findings from exploring the DiaNAH dataset could have significant implications for the assessment and management of visual impairments in ABI patients, ultimately improving their quality of life.

This study represents an initial effort to systematically explore and preprocess the DiaNAH dataset in order to enable further exploration of the complex interplay between visual complaints and visual perceptual function in ABI patients, paving the way for future research in this area.

\section*{Methodology}
The DiaNAH dataset comprises patients who suffer from ABI, each of which completed a computer vision syndrome questionnaire (CVS) questionnaire and a visual impairment test battery. The CVS questionnaire is a tool utilized to gather patient data on phenotypes, encompassing both symptoms and causes.
The total number of ABI patients within the DiaNAH dataset is 948. However, due to a range of exclusion criteria that resulted in missing values (further elaborated for each test under \textit{Appendix - Neuropsychological Tests and Evaluations}), some patients were precluded from the analysis. In total, 181 patients were omitted, resulting in an analytical dataset comprised of 767 patients (an in-depth description follows in Subsection \textit{Missing Values}).
In the context of this study, complaints were defined as patient responses to specific questions about their general visual health. The subsequent table illustrates these complaints:

\begin{table}[H]
\centering
\begin{tabular}{|c|p{0.7\linewidth}|}
\hline
\textbf{Complaint question} & \textbf{Question} \\ [0.5ex] 
\hline\hline
CVS 12 & Do you have blurred vision? \\ 
\hline
CVS 13 & Do you have trouble seeing letters in a text against the background? \\
\hline
CVS 14 & Are you blinded by bright light more than before the brain injury? \\
\hline
CVS 15 & Do you have the impression that everything seems darker (do you experience a dark haze) than before the brain injury? \\
\hline
CVS 16 & Do you need more light when you see than before the brain injury? \\
\hline
CVS 18 & If you go out into bright sunlight, does it take longer than before the brain injury to get used to the bright light? \\
\hline
CVS 19 & When you go back inside from the bright sunlight, does it take longer than before the brain injury to get used to it? \\
\hline
CVS 20 & Do you experience colors differently than before the brain injury? \\
\hline
\end{tabular}
\caption{Summary of Complaint Questions Used in the Survey.}
\label{table:1}
\end{table}

\begin{table}[H]
\centering
\begin{tabular}{|c|c|c|}
\hline
\textbf{Complaint question} & \textbf{Description} & \textbf{Label} \\ [0.5ex] 
\hline\hline
CVS 12 & Blurred vision & 1000 0000 \\ 
\hline
CVS 13 & Contrast perception & 0100 0000 \\
\hline
CVS 14 & Light nuisance & 0010 0000 \\
\hline
CVS 15 & Dark haze & 0001 0000 \\
\hline
CVS 16 & Increased need of light & 0000 1000 \\
\hline
CVS 18 & Disrupted light adaptation & 0000 0100 \\
\hline
CVS 19 & Disrupted dark adaptation & 0000 0010 \\
\hline
CVS 20 & Color differences & 0000 0001 \\
\hline
\end{tabular}
\caption{Descriptions and Labels of Complaint Questions}
\label{table:2}
\end{table}

\subsection*{Correlation}
In terms of features, the DiaNAH dataset contains a total of 389 variables. When deploying Machine Learning algorithms such as clustering, data with many variables (high-dimensionality) can be undesirable, so it is important to select only the variables that look promising. For this, a cross-variable correlation test is conducted and displayed in Figure ~\ref{fig:heatmap} (see \ref{appendix:corrMatrix}).

To visualize the correlation matrix, a heatmap was used:

\begin{figure}[H]
    \centering
    \includegraphics[width=0.9\columnwidth]{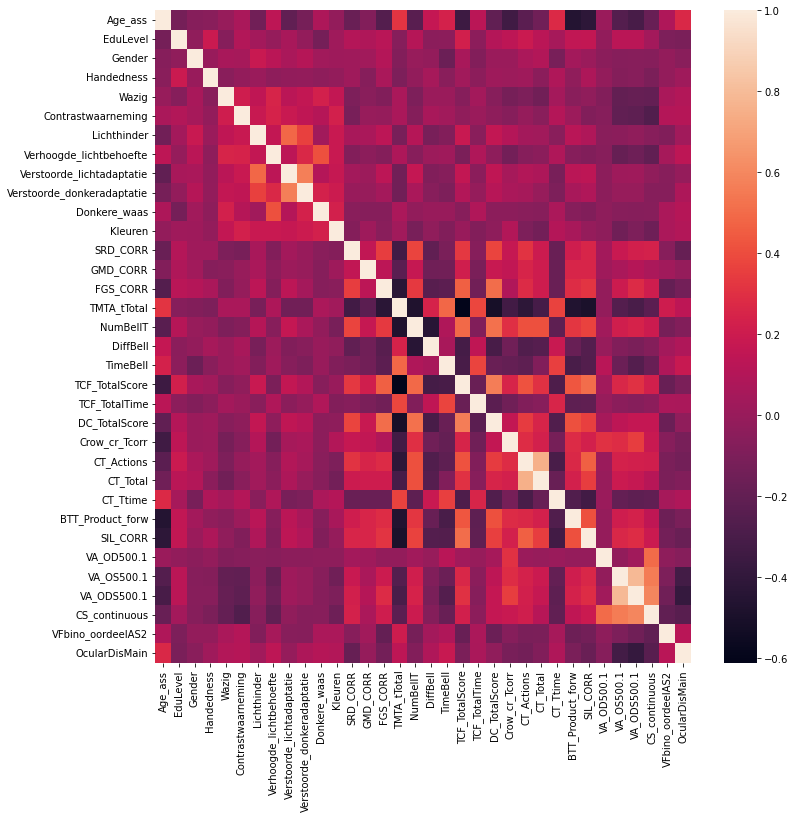}
    \caption{Visualization of the Correlations Between the Complaints and Visual Perceptual Functions.} 
    \label{fig:heatmap}    
\end{figure}

Figure ~\ref{fig:heatmap} shows positive and negative correlations may be distinguished in this way: the lighter the color the more positively correlated two variables are (if one moves in the positive direction the other one follows in the positive direction), the darker the color the more negatively correlated two variables are (if one moves in the positive direction the other one follows in the negative direction).

\begin{figure}[H]
    \centering
    \includegraphics[width=1\columnwidth]{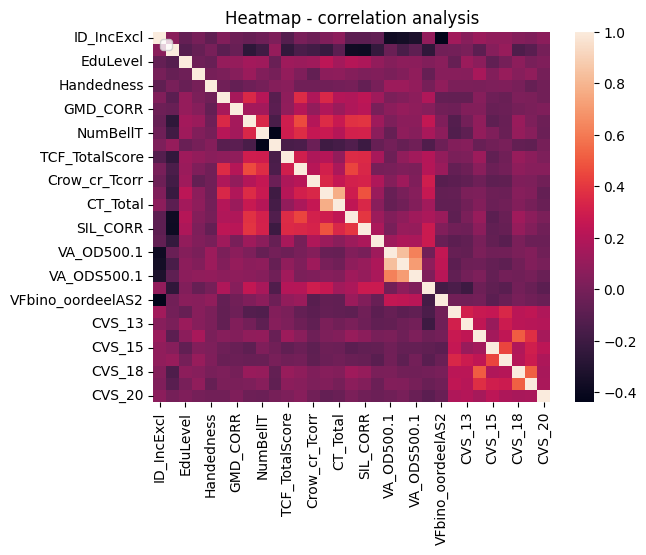}
    \caption{Visualization of the Correlations Between the Complaints and Visual Perceptual Functions of the Filtered Features.}
    \label{fig:heatmap_filtered}    
\end{figure}

We must remember that correlation does not imply causation, and a high absolute correlation could be due to confounding variables or other factors. As such, some complaints correlate strongly with each other, and there are also high correlations between visual perceptual function tests. However, the complaints are only moderately or not at all correlated with the perceptual functions ($\approx 0.0$). Therefore, there seems to be no clear connection between these complaints and visual perceptual function tests using this linear correlation method. In Figure ~\ref{fig:heatmap_filtered} we have another heatmap which resulted from removing variables that only showed moderate to no correlation across the dataset.

\subsection*{Missing values}\label{subsec:missingvalues}

The DiaNAH dataset has a subset of the original patients: those with ABI which have completed the CVS questionnaire.
The dataset then consists of 948 patients with ABI. Out of these 948 patients, 181 are excluded from the analysis due to unreliable performance during the test battery. 

\begin{figure}[H]
    \centering
    \includegraphics[width=1\columnwidth]{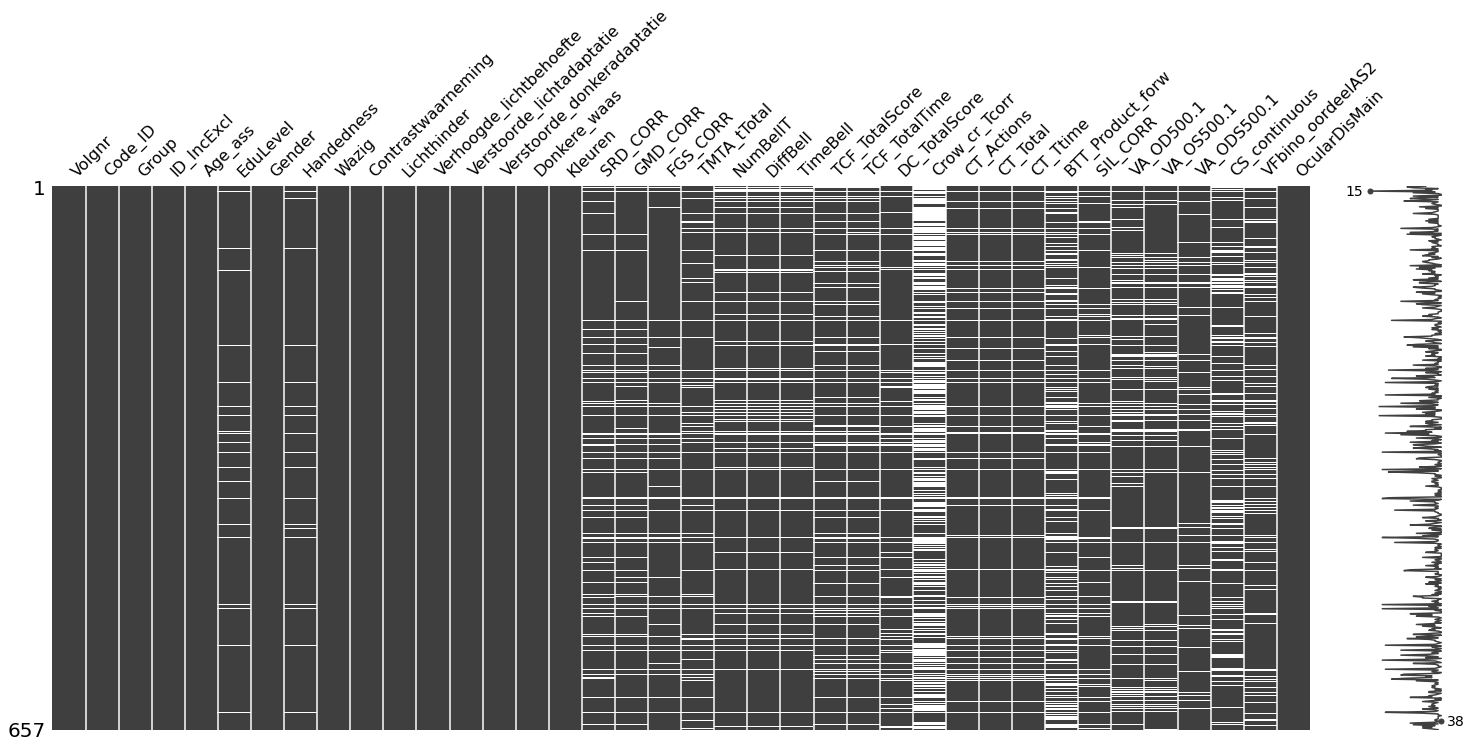}
    \caption{Overview of the Missing Values Within the DiaNAH Dataset.}
    \label{fig:missingvalues}   
\end{figure}

The above matrix shows whether a feature has a value (black) or whether it is a missing value (white) per patient. The patients are displayed below each other in rows, while the variables (n=38) are displayed in columns. On the right is a graph in which the total number of variables per patient has a value, with the minimum being one patient with only 15 variables, and the maximum one with all 38 variables. It becomes evident, for example, that Crow\_cr\_Tcorr, the test outcome measure of the \textit{Crowding test} (see ~\ref{appendix:visualtests}), has many missing values when compared to other variables. 

\begin{figure}[H]
    \centering
    \includegraphics[width=1\columnwidth]{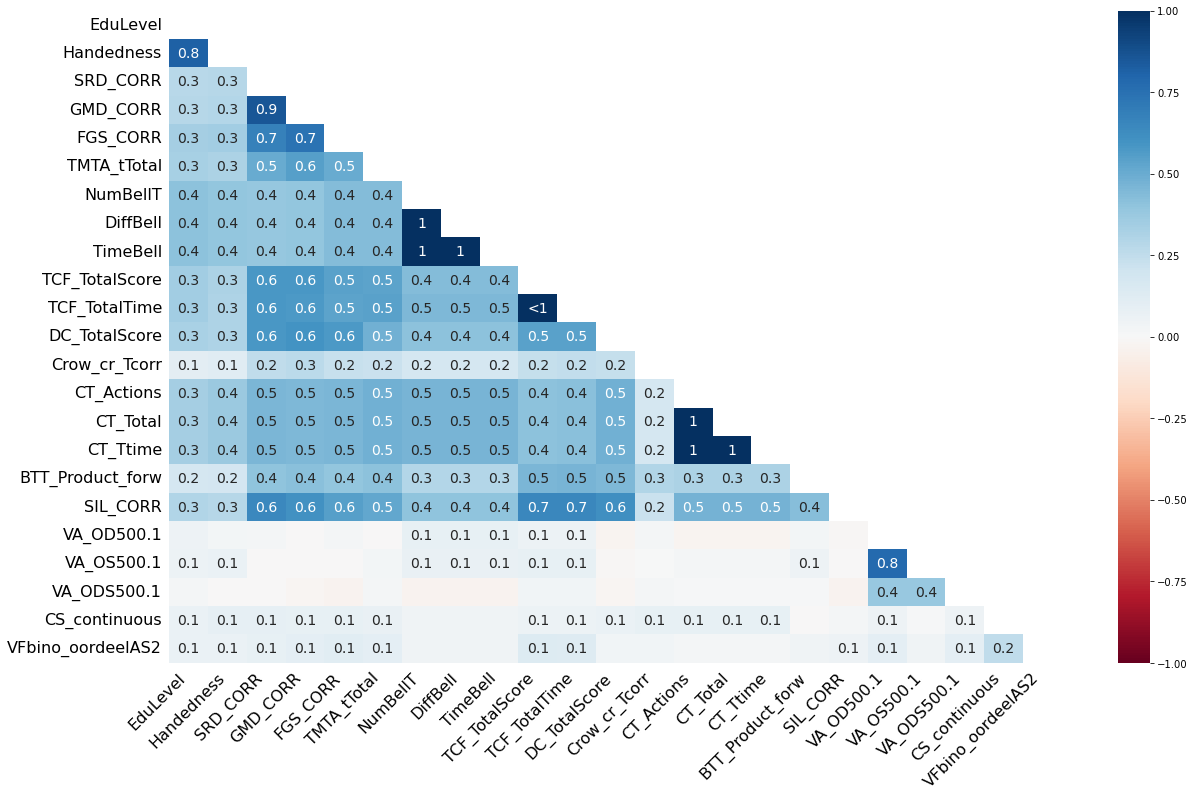}
    \caption{Overview of Correlations Between Test Outcome Measures Based on Missing Data.}
    \label{fig:correlation2}   
\end{figure}

About 5-10\% per variable data is missing and this missing data is non-random according to the domain experts, being due to some tests not having been taken or finished. Some possible reasons reported by the experts are that there was not enough time for a given patient to take all the tests, the tests were too difficult for the patient, or not at all possible for the patient to perform. In addition, a test may have been aborted by the test performer for other reasons (e.g. a patient had to perform the tests so slowly that it rendered the result unreliable). It is therefore important to evaluate whether there is a pattern in the missing values and how they can possibly be supplemented and incorporated into the model, using the correlation matrix in Figure ~\ref{fig:correlation2}. 

Figure \ref{fig:correlation2} provides an analysis of correlation between missing values of the selected variables. A high value (maximum 1) in the matrix shows a correlation of missing values, where if one is missing the other is also likely to be missing, while a value around 0 shows no correlation. Test outcome measures from the same test such as CT\_actions, CT\_total, CT\_Ttime therefore have a value of 1. This analysis was then used during imputation as an exclusion factor for patients in the following way. Given that a patient is missing all variables pertaining to a given test, it must be excluded since this test cannot be predicted from the other with which it correlates. In other words, if a patient is missing the time variable of a test but has the score variable, the imputation model can still make a prediction for the time variable given the score variable. If a patient misses both variables, where a correlation of 1 exists between two variables within the same test, then this prediction is far too weak to be considered.

It is important to note that different patients can have one or multiple complaints, thus the groupings of complaints are complex. The following is a histogram of representative complaints in the dataset (where five or more people share the same complaint group). 

\begin{figure}[H]
    \centering
    \includegraphics[width=0.8\columnwidth]{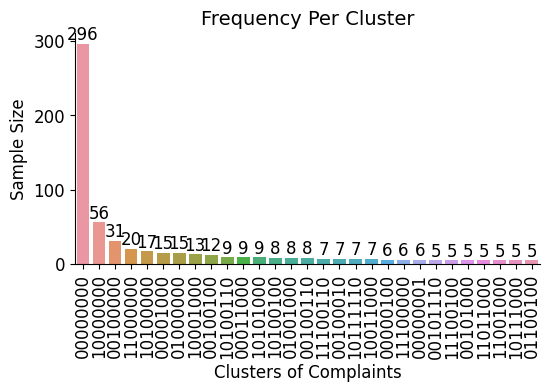}
    \caption{Histogram of Frequency per Clusters of Complaints.}
    \label{fig:histogram_complaints}   
\end{figure}

The above histogram, Figure ~\ref{fig:histogram_complaints}, shows complaints and complaint combinations across patients. That is, out of all the available combinations of complaints, some patients have singular complaints while others have multiple complaints. Below is a table showcasing the available singular complaint groups:

\begin{table}[H]
\centering
\scriptsize
\begin{tabular}{|c|c|c|c|}
\hline
\textbf{Label} & \textbf{Code} & \textbf{CVS question} & \textbf{Sample Size} \\ [0.5ex] 
\hline\hline
No Complaints & df\_00000000 & CVS N/A & 296 \\ 
\hline
Color diff & df\_00000001 & CVS 20 & 6 \\
\hline
Light adaptation & df\_00000100 & CVS 18 & 6 \\
\hline
Increased need light & df\_00001000 & CVS 16 & 15 \\
\hline
Light nuisance & df\_00100000 & CVS 14 & 31 \\
\hline
Contrast perception & df\_01000000 & CVS 13 & 15 \\
\hline
Blurred vision & df\_10000000 & CVS 12 & 56 \\
\hline
\end{tabular}
\caption{Singular Complaint Groups.}
\label{table:3}
\end{table}

The sample size in our study is small, making certain aspects unknowable due to the explosion of possible combinations. In theory, there can be $8! = 40,320$ possible groupings of complaints - assuming each complaint is not taken to be a gradient or scale, otherwise the number of possible complaint combinations is even larger, probably by several orders of magnitude. Out of these, the data gives access to only 30 complaint groups, with at least 5 patients each, while missing a representative group of individual complaints for CVS\_15 and CVS\_19.

\subsubsection*{Power Analysis}

A power analysis is important to gauge the relevant sample size given some underlying assumptions (see \ref{appendix:poweranalysis}).
We conducted a power analysis assuming a large effect size of $0.8$ given the strongly held assumptions that the tests are testing what they are supposed to, a significance level of $0.05$, and a desired statistical power of $0.8$. 
If we take a clustering or classification analysis that is supposed to differentiate between two groups, one with no complaints (\#1) and another with complaints (\#2), the sample sizes for groups \#1 and \#2 were calculated as per \ref{appendix:poweranalysis}:

\begin{table}[H]
\centering
\begin{tabular}{|c|c|}
\hline
\textbf{Sample Group} & \textbf{Size} \\ 
\hline\hline
Group 1 & 14 \\ 
\hline
Group 2 & 14 \\
\hline
Total & 28 \\
\hline
\end{tabular}
\caption{Sample Sizes.}
\label{table:SampleSizes}
\end{table}

\begin{table}[H]
\centering
\begin{tabular}{|c|c|}
\hline
\textbf{Parameter} & \textbf{Value} \\ 
\hline\hline
Incidence, group 1 & 75\% \\ 
\hline
Incidence, group 2 & 25\% \\
\hline
Alpha & 0.05 \\
\hline
Beta & 0.2 \\
\hline
Power & 0.8 \\
\hline
\end{tabular}
\caption{Study Parameters.}
\label{table:StudyParameters}
\end{table}

Table ~\ref{table:SampleSizes} lists the sample sizes for each group and the subsequent desirable total. The second table (Table \ref{table:StudyParameters}) lists the study parameters: the incidence for each group, and the values of alpha, beta, and power given the present underlying assumptions.

The above analysis indicates that, given these parameters, the current sample size may limit our ability to detect smaller effects since at best, looking at the largest complaint group, we have a $61:61$ split in the case of patients with blurred vision and patients with no complaints, and at worst we have $5:5$ splits. Despite of the results of the power analysis and due to the present complaint groups sample sizes, we decided to pursue groups with at least 7 patients. This allows us to better establish our methodology because more complaint group comparisons yields more experiments. For this reason results should be interpreted with caution, and further studies with larger sample sizes are recommended to more reliably detect effects and explore the variety of potential complaint groupings.

\subsection*{Data Imputation}
The process began with feature categorization, in which features were numerically categorized where possible. Informed by the correlation analysis in Figure \ref{fig:heatmap_filtered}, features containing demographic information such as education level, handedness, and gender were omitted due to their lack of direct relevance to visual complaints. Subsequently, one-hot encoding (the process of encoding or converting categorical data into a binary vector where only one element is "hot", or 1, and the rest are "cold", or 0) of the complaint labels were generated to facilitate the deployment of machine learning algorithms.

In order to bolster the robustness of our analysis, two separate datasets were constructed. The first dataset was assembled around patient clusters displaying \textbf{singular complaints}, totalling 129 patients, that is, patient groups with only one singular complaint each. The second dataset comprised of patients within any complaint clusters, encompassing a total of 767 patients.

Feature availability was a critical determinant of patient inclusion in our datasets. As mentioned earlier, patients were excluded based on specific exclusion criteria per test. We also dropped all entries that missed half the data (totalling 30 patients). Then deleted patients that were missing all evaluations of any given test (were missing an entire test): removed 28 datapoints that simultaneously missed the data under NumBellT, DiffBell, and TimeBell; and another 27 datapoints that simultaneously missed the data under CT\_Actions, CT\_Total, and CT\_Ttime. This is so that the AutoML (automated machine learning) method can reliably fill in missing data for a single datapoint basing the missing distributions on in-between test comparisons. The following is a table showing the missing data per test result.

\begin{table}[H]
\centering
\begin{tabular}{lr}
\hline
\textbf{Variable}         & \textbf{Missing Values} \\ \hline
SRD\_CORR              & 94  \\
GMD\_CORR              & 89  \\
FGS\_CORR              & 60  \\
TMTA\_tTotal           & 72  \\
NumBellT              & 54  \\
DiffBell              & 54  \\
TimeBell              & 54  \\
TCF\_TotalScore       & 117 \\
TCF\_TotalTime        & 118 \\
DC\_TotalScore         & 84  \\
Crow\_cr\_Tcorr        & 230 \\
CT\_Actions            & 56  \\
CT\_Total              & 56  \\
CT\_Ttime              & 55  \\
BTT\_Product\_forw     & 148 \\
SIL\_CORR              & 88  \\
OcularDisMain          & 0   \\
CS\_continuous        & 191 \\
VFbino\_oordeelAS2    & 137 \\ \hline
\end{tabular}
\caption{Missing values across data}
\label{tab:missing_values}
\end{table}

To handle these incomplete features, an automated machine learning method was implemented utilizing PyCaret, an AutoML library coded in Python. The analysis relied on two specialized model batteries: one focusing on categorical features and the other on continuous features. What this method enables is a strict convergence between imputed values and the known sample distributions. By using the automatically fitted models the proceeding generated values should fall under the desirable distributions found in the available data. In other words, the imputed values are generated from a holistic distribution that considers every available variable with varying weights of importance, automatically determined by the model battery itself.

First, the two datasets, continuous and categorical, were each normalized.

Given a dataset \( D \) with incomplete features, the imputation process can be mathematically described as follows:

Let \( D = \{x_1, x_2, \ldots, x_n\} \) be the dataset where each \( x_i \) represents a feature vector.

We assume that some of the features in \( D \) are incomplete. We denote the incomplete features as $D_{\text{incomplete}}$ and the complete features as $D_{\text{complete}}$.

\begin{align}
   D_{\text{incomplete}} \subset D \ &&
   D_{\text{complete}} = D \setminus D_{\text{incomplete}}
\end{align}

The imputation process using models can be described in the following steps:
\begin{enumerate}
  \item \textbf{Feature Categorization}: Divide features into categorical (e.g. gender) and continuous (e.g. time taken at a test) subsets, denoted as \( D_{\text{cat}} \) and \( D_{\text{cont}} \) respectively.
  \item \textbf{Model Training}: Train two separate models, one for each type of feature:
  \begin{itemize}
    \item Model \( M_{\text{cat}} \) for categorical features \( D_{\text{cat}} \).
    \item Model \( M_{\text{cont}} \) for continuous features \( D_{\text{cont}} \).
  \end{itemize}
  \item \textbf{Imputation}: For each incomplete feature \( x \in D_{\text{incomplete}} \), perform the following:
  \begin{itemize}
    \item If \( x \) is categorical, use \( M_{\text{cat}} \) to predict the missing values.
    \item If \( x \) is continuous, use \( M_{\text{cont}} \) to predict the missing values.
  \end{itemize}
  \item \textbf{Integration}: Combine the imputed values with \( D_{\text{complete}} \) to form the fully imputed dataset \( D_{\text{imputed}} \).
\end{enumerate}

The goal is to ensure that the imputed values closely match the known distributions of the available data in \( D \). This is achieved by using the models \( M_{\text{cat}} \) and \( M_{\text{cont}} \), which are trained to capture the underlying patterns and distributions of the respective feature types.

\subsection*{Model Training}
We trained two models: \( M_{\text{cat}} \) for \( D_{\text{cat}} \) and \( M_{\text{cont}} \) for \( D_{\text{cont}} \). These models learn the distributions \( P(x) \) for each feature type.

\subsection*{Imputation Process}
Imputation for a missing feature \( x \) is modelled as:
\[ x_{\text{imputed}} \sim P(x | D_{\text{complete}}) \]
where \( D_{\text{complete}} \) represents the subset of \( D \) without missing values. This process aligns the imputed values with the statistical characteristics of the observed data.

\subsection*{Model Efficacy in Imputation}
We posit that a model-based imputation is superior to average imputation. In average imputation, the model predicts a constant value equivalent to the mean of the observed data which can be skewed by outliers or extreme values:
\[ f(x) = \bar{y}, \forall x \in X \]
While in a model-based approach, the model leverages all available data, offering predictions that should reflect the conditional expectations and the data's statistical intricacies:
\[ f(x) \approx E[Y | X=x] \]

\subsection*{Utilizing Data Distributions}
\begin{figure}[H]
    \centering
    \includegraphics[width=1\columnwidth]{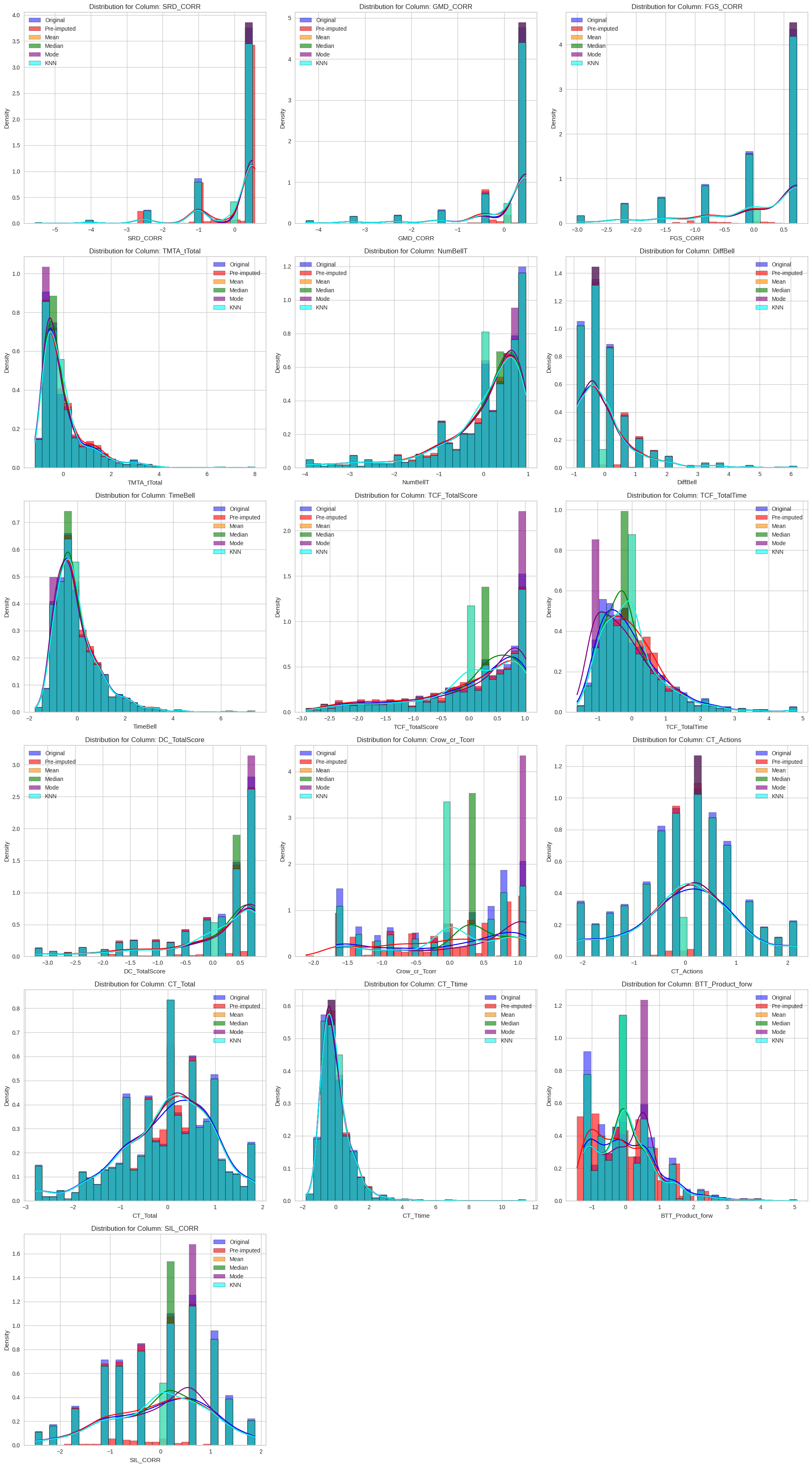}
    \caption{Histogram comparing the imputation methods.}
    \label{fig:histogram_impute_methods}   
\end{figure}

The models \( M_{\text{cat}} \) and \( M_{\text{cont}} \) are designed to approximate the conditional probability distributions \( P(y | x) \), imputing values that maintain the dataset's original statistical properties. This approach is more useful than simple averaging, as it takes into account the relationships and patterns inherent in the data (see Figure ~\ref{fig:histogram_impute_methods}.

\section*{Results}
Firstly, addressing missing values, the white spaces in the non-visualized matrix (Figure \ref{fig:missingvalues}) highlighted a relatively frequent absence of data. For the more robust application of machine learning, PyCaret's AutoML was deployed to compute imputations for the data. Thus, the generated imputations should resemble the known data distributions closely, adhering to naturally arising patterns within the data while acknowledging the possible non-random nature of the missing values.

In order the gauge the proposed imputation method with standard methods, we compare the results. In the context of test performance metrics for people with visual disabilities, it's important to consider which statistical measures are most critical for this analysis. Since the data is skewed and the distribution is non-standard, certain metrics like median, skewness, and kurtosis are more relevant than others. Therefore, we propose the following rationale for a weighted evaluation vector and the corresponding weight (w):
\begin{enumerate}[(a)]
\item Mean: It is less representative given that the data is highly skewed ($w=0.10$).
\item Standard Deviation (std): Important for understanding variability in test performance ($w=0.10$).
\item Median: Relevant in skewed distributions as it gives a better sense of the central tendency ($w=0.15$).
\item Minimum (min) and Maximum (max): Less critical but provides insights into the range of performances ($w=0.05$).
\item Skewness: Very important, as it will indicate how symmetrical the data is, which is crucial in skewed distributions ($w=0.30$).
\item Kurtosis: Important to understand the `tailedness' of the distribution, which can be relevant for understanding extremes in performance ($w=0.30$).
\end{enumerate}

These weights sum up to 1. The median and skewness are given more weight because of the skewed nature of the data and the focus on central tendency in such distributions.

We then imputed the data using the various methods listed before and the AutoML method. Below are the results:

\begin{table}[H]
\centering
\begin{tabular}{|l|c|}
\hline
\textbf{Imputation Method} & \textbf{Average Weighted Deviation} \\
\hline
ML-solution                & 0.002384                            \\
Mean                       & 0.002015                            \\
Median                     & 0.001179                            \\
Mode                       & 0.002690                            \\
KNN                        & 0.002015                            \\
\hline
\end{tabular}
\caption{Average Weighted Deviation for Different Imputation Methods}
\label{tab:imputation_methods_deviation}
\end{table}

When evaluating the effectiveness of the various imputation methods, we consider the specific characteristics and requirements of the dataset. In the case of the provided data, the weighted vector places substantial emphasis on skewness and kurtosis, each accounting for 30\% of the weight. This emphasis underscores the importance of preserving the original distribution of the data, particularly its shape and tail behavior.

Analyzing the average weighted deviations of different imputation methods, we observe that traditional methods like Mean, Median, and KNN yield lower deviations compared to the ML-based pre-imputation method, with Median imputation performing the best. This implies that traditional methods are superior. But, as was stated before, ML-based imputation techniques have a distinct advantage that is not captured in the weighted average, in modelling more intricate relationships within data, as they consider multiple features or variables simultaneously. This capability is particularly important in datasets where missingness is not random and may depend on other variables. We assume this to be the case in the DiaNAH dataset, therefore, while traditional methods might minimize deviations in standard statistical measurements, they might not capture the full complexity of the data's distribution, particularly in how different features correlate with one another. In the following plot, "pre-imputed" refers to the introduced AutoML method. The AutoML method predicts a continuous feature  distribution over the predicted features. The predicted values also closely match the original distribution. Particularly, it does not overshoot or undershoot the possible scores, while following the skewness as well as the hump of the feature's distributions.

Our correlation matrix and heatmap visualization (Figure \ref{fig:heatmap_filtered}) suggested interesting patterns of association among certain variables. Particularly, there was a notable degree of intercorrelation between the complaint variables, yet only moderate to low correlation with the visual perceptual function test variables, suggesting a possible dissociation between perceived visual deficits and measured visual abilities.

Certain complaint groups are considerably more populous than others, as noted. However, the wide variation in complaint group sizes and the vast potential for various group combinations (up to 40,320) means that the largest majority of possible complaint groupings are not represented within the available data. In fact, the data represents $0.07\%$ of the 40k possible groups. This fact alone poses many questions regarding the prevalence and nature of visual complaints post-ABI and what the research community can really know about them.

These results should catalyze further research, especially considering the recognized limitations such as the definite underrepresentation of most complaint groupings and the notable presence of missing data.

\section*{Discussion}
Our exploration revealed numerous challenges and complexities entwined within the dataset, particularly in regard to the number of potential complaint combinations amongst the ABI patients. As laid out, with the theoretical possibility of $8! = 40,320$ distinct complaint combinations, there arises a challenge to sufficiently understand and subsequently classify patient complaints in a methodical and clinically useful manner. Even within our substantive dataset, only a small number of these combinations were observed ($\approx 0.07\%$), illustrating the critical need for even more extensive data collection and analysis to navigate through the labyrinth of patient experiences in visual complaints after ABI.

Our findings, notably regarding correlations among different complaints and visual perceptual function tests warrant cautious interpretation. The absence of a linear correlation (Figure ~\ref{fig:heatmap}) does not eradicate the possibility of an existing relationship; it merely points towards the non-linear, potentially multifaceted nature of such a relationship. Therefore, future endeavors will explore the potential utility of non-linear models and machine learning algorithms which can unravel these complex relationships in a more nuanced manner.

In the context of clinical application, these findings, albeit initial, signal towards the potential existence of distinctive visual complaint profiles amongst ABI patients. However, the limited sample size and the skewed distribution across various complaint groupings post a noteworthy limitation to the generalizability and applicability of the findings. It is therefore important that future research scales these explorations to a more expansive dataset, ensuring a wider representation of complaint combinations and subsequently, more clinically-relevant categorizations.

We consider our study into patient experiences provided by analyzing their visual complaints to hold valuable implications for tailoring patient-centric intervention strategies post-ABI. Visual complaints could potentially be deployed as markers, guiding clinicians in deciphering the unique, individualized impacts of ABI on visual functioning and quality of life for each patient. Future work should thus pivot towards not only amplifying the robustness of these initial findings but also venturing to integrate them into tangible, impactful clinical applications, thereby bridging the divide between research findings and applied clinical practice.

In conclusion, we have prepared a dataset that can be useful in testing many algorithms, like clustering or classification. We have also observed a disconnect between visual perceptual function tests and complaints, laying a foundation for future studies to explore the non-linear intricacies among these variables, thereby enhancing our understanding and ultimately furthering our approach towards addressing visual complaints and improving the quality of life for ABI patients.

\newpage
\section{Neuropsychological Tests and Evaluations}
\label{appendix:visualtests}
The DiaNAH dataset consists of test scores of multiple tests. In this section, a description is given for each relevant test, namely regarding the tasks, followed by a discussion of possible checks that can be implemented to confirm a good measurement. In essence this section contains the rationale for preprocessing decisions of the data.

\subsection*{Spontaneous drawing test}
At the begining of the test battery, tests are done to accustom subjects to the tablet and pencil stylus. These tests are also done to gauge the subject's disorder severity in how it affects hand movement. The subjects were asked to produce one or several of the options below:
\begin{itemize}
\item Write a sentence
\item A clock (at 10 past 11)
\item A 5-point star
\item A solid cube in perspective
\end{itemize}

These tasks can give quick insights into possible motor/mental disabilities besides visual impairments. Disabilities (especially motor impairments) will increase the time per task for all pen tests or even make it impossible for the subject to perform the test. The tasks that use a pen are the Bell's test, Trail making test, Corsi Block-tapping test and the Taylor Complex figure test.

\subsection*{Bell's test}
The Bells Test, \cite{Gauthier1989}, is a cancellation task for peripersonal visual neglect, assessing visual search strategies and selective attention. Participants cross out target stimuli (bells) among distractors, with outcomes including total number of bells crossed and time taken.

\begin{figure}[H]
   \centering
   \includegraphics [width=0.7\textwidth] {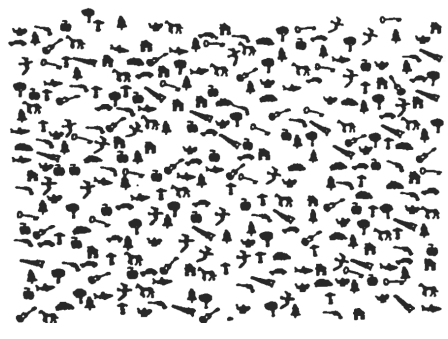}
  \caption{Bells Test task sheet presented to the subject.}
\end{figure}

For this task the pen is used on the A3 scoring board. Subjects with selective attention, reduced information processing speed, visual field deficit and loss of executive control will all likely take longer to finish this task than the "average" person of their age.

\subsubsection*{Exclusion considerations}
\begin{itemize}
    \item Cut-offs: based on literature (Iris Tigchelaar) a recommended cut-off of 6 bell's and 5 minutes should exclude respectively 5\% to 12\%  of healthy test subjects.
    \item Outliers: Subjects with extreme perception impairments are the outliers.
    \item Scores and Time: Include any subject who has a score AND time. Even if the score is 0 and only a few seconds are recorded, as it is unclear what happened in that time
    \item Insight (InsightBell): If a score is entered it is reasonable to expect the test to be stopped prematurely. Based on previous pilot this seems reasonable and does not skew data.
\end{itemize}

\subsubsection*{Outcome variables}
"NumBellT", "DiffBell", "TimeBell".

\subsection*{Cake Thief/Birthday party test}
The Birthday Party Test (University of Groningen; department of developmental and clinical neuropsychology) is used to screen for simultanagnosia. It consists of a complex image of a scene where a typical situation encountered in everyday life takes place \cite{de2022birthday}. The picture is a two-dimensional line drawing in A4-landscape-format depicted in black lines on a white background for maximal contrast. The participant is instructed to describe out loud, as accurately and thoroughly as possible, what objects can be seen and what actions are taking place in the picture. The participant is encouraged by the experimenter to provide more information in case only details of the picture are mentioned and if those details are not integrated into a coherent scene, or if the participant is not giving further responses.

\begin{figure}[H]
   \centering
   \includegraphics [width=0.7\textwidth] {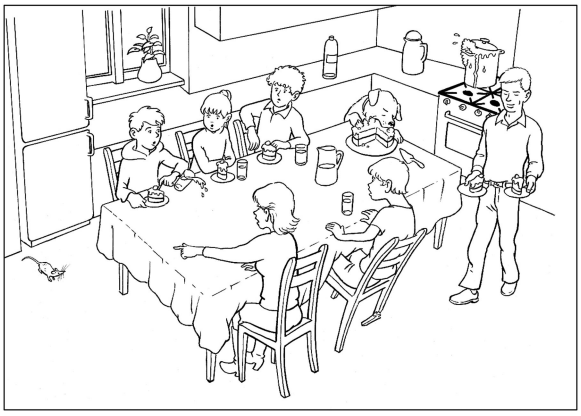}
  \caption{Birthday party test task sheet presented to the subject.}
\end{figure}

Scoring is facilitated by a scoring aid that presents a complete list of items that are depicted on the Birthday Party Test. 

\subsubsection*{Outcome variables}
 "CT Actions", "CT Total", "CT Ttime".

\subsection*{Trail making test}
The Trail Making Test \cite{reiten} is a widely used visuomotor task that is thought to rely on different cognitive functions, such as attention, visual search, and processing speed \cite{macpherson, strauss}. In part A, the participant has to connect in ascending order 25 numbers that are distributed on an A4-sized portrait-format. In part B, the participant is instructed to connect numbers and letters by alternating them (1-A … L-13). The ratio score (B/A-index: time B divided by time A) is calculated as a measure of cognitive flexibility corrected for visual motor speed. The time is automatically registered from the moment the pen hits the tablet until the last item that is connected by the participant.

\begin{figure}[H]
   \centering
   \includegraphics [width=0.7\textwidth] {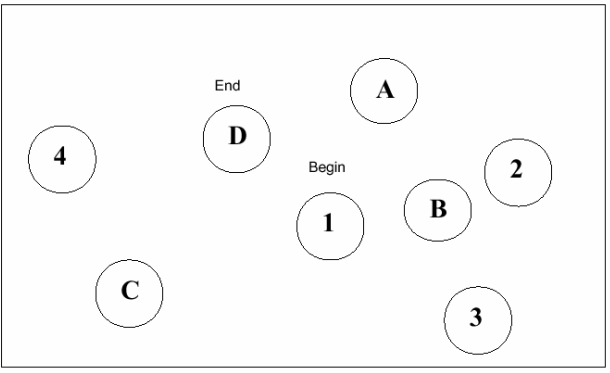}
  \caption{Trail making part B example presented to the subject.}
\end{figure}

If patients get it wrong, they are told they made a mistake (they have to finish the test without having made a mistake) - in that case, they backtrack one step. If people take 10 min the test is stopped. B|A result measures mentally flexibility : A is number to number and B is number to letter.

\subsubsection*{Outcome variables}
"TMTA tTotal".

\subsection*{L-POST: Figure Ground Segmentation, Global Motion Detection, Shape Ratio Discrimination}
Three of the L-POST tests, \cite{Torfs2013TheLP} ,were selected for the DiaNAH test battery, including Figure Ground Segmentation (FGS), Global Motion Detection (GMD) and Shape Ratio Discrimination (SRD). The test follows a matching-to-sample procedure, where the target needs to be matched to one of three alternatives e.g. multiple-choice. Each of the tests is preceded with two example items (except for SRD), consists of five items and is self-paced. A subject is asked to always give an answer even if it is a guess. The three tests are the following:

\begin{itemize}
    \item FGS is thought to assess the ability to discriminate figure from background.
    \item GMD is thought to assess the ability of the participant to recognise a group of coherently moving dots between randomly moving dots.
    \item SRD is thought to assess visual form recognition of black objects on a white background.
\end{itemize}

\subsubsection*{Exclusion considerations}
\begin{itemize}
    \item SRD\_Completed: If the test is completed the data is included and if not, then it is completed excluded.
    \item FGS\_Completed: The same in-/exclusion criteria as SRD.
    \item GMD\_Completed: The same in-/exclusion criteria as SRD.
\end{itemize}

\subsubsection*{Outcome variables}
"SRD CORR", "GMD CORR", "FGS CORR".

\subsection*{Corsi Block-tapping test}
The Corsi Block-Tapping Test is a widely used task for visuospatial memory \cite{milner, corsi}, in which a participant is required to repeat a sequence of movements by tapping blocks. In the digitized version of this task, a sequence of blocks flash yellow (r,g,b = 255,220,0) for 500 ms with an interblock interval of 1000 ms in a certain order. The participant is instructed to repeat the blocks in the same order by tapping on them. The test is automatically discontinued if two trials of the same length are incorrectly repeated by the participant.

\begin{figure}[H]
   \centering
   \includegraphics [width=0.7\textwidth] {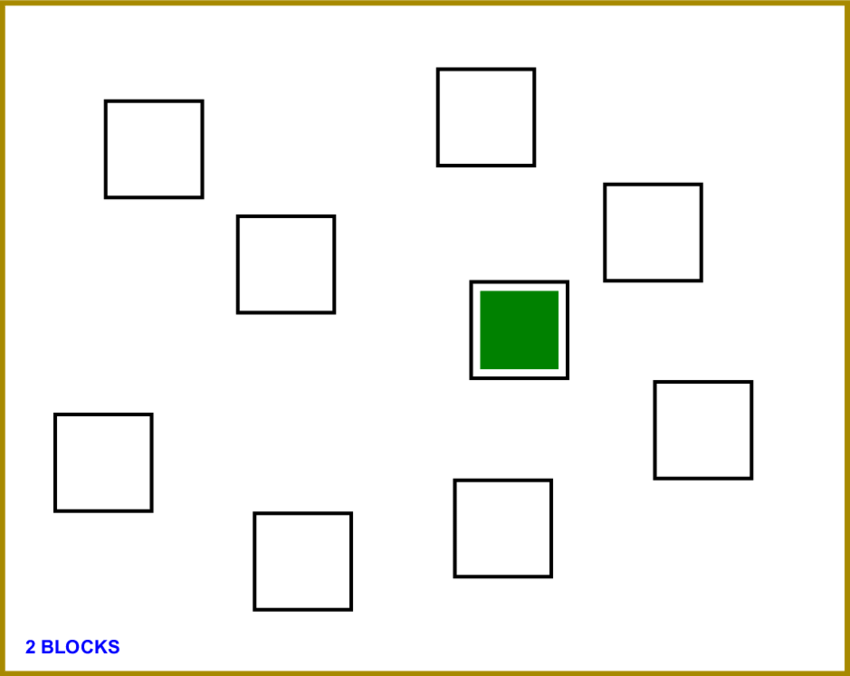}
  \caption{A slightly different version of the Corsi-blocks as presented to the subject.}
\end{figure}

\subsubsection*{Exclusion considerations}
\begin{itemize}
    \item BTT\_incl == 0: Exclusion of all participants with a total score of 0. However it should be noted that this is likely to exclude a portion of the participants that we would want to detect with the Corsi test (e.g. severly visuospatial memory impairments).
    \item BTT\_incl $>= 1$: Include all participants with a score of 1 or above.
    \item BTT\_incl == NA: missing test means it was not started. Exclusion is needed.
    \item BTT\_ExclReason == Any: If an exclusion reason was given the participant is excluded. Some reasons could be:
    \begin{itemize}
        \item 1 Participant has performed the test before in a different context.
        \item 2 Malingering study: participant indicates not to have made this test.
        \item 3 Data unreliable because of technical problems, like test disruption.
        \item 4 BTT Tcorr forw == 0 (same as first item in list).
    \end{itemize}
\end{itemize}

\subsubsection*{Outcome variables}
"BTT Product forw".

\subsection*{Dot counting test}
The Dot Counting Test aims to measure non lateralized spatial attention, i.e. visual search/grouping, and is derived from the Dot Counting subtest of the Visual Object and Space Perception battery \cite{vosp} and from the dot counting subtest of the L-POST \cite{vancleef}. 

An A5-portrait size image containing between five and ten black dots with a diameter of five mm is presented on the screen and the participant has to indicate the number of dots that are depicted. The dots disappear after 5 seconds, making serial counting more difficult. However, this gives patients with, for example, visual field defects (adequate compensatory behavior assumed) sufficient time to acquire an overview of the image.

\subsubsection*{Exclusion considerations}
\begin{itemize}
    \item DC Completed: If the test is finished (e.g. True) the data is included. Data is excluded if 0 or NA value for this feature.
\end{itemize}

\subsubsection*{Outcome variables}
"DC TotalScore".

\subsection*{Figure taylor}
The Taylor Complex Figure is designed to evaluate visuoconstructive abilities and long-term memory \cite{lezak,taylor}. The original research contains three conditions: copy, immediate recall, and delayed recall. For the DiaNAH test battery, only the copy test was included as the aim is to measure visuoconstructive functions.

\begin{figure}[H]
   \centering
   \includegraphics [width=0.5\textwidth] {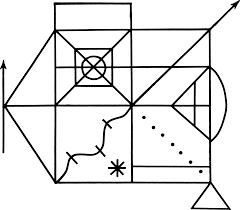}
  \caption{Tayor Complex figure as presented to the subject.}
\end{figure}

The participant is instructed to copy the Taylor Complex Figure as accurately as possible. The figure is fractionated into 18 different elements. For each element, a maximum score of two points is allocated if correctly drawn and correctly placed, one point if correctly drawn, but incorrectly placed, one point if distorted/incomplete but correctly placed, 0.5 points if distorted/incomplete but recognizable and incorrectly placed. No points are awarded if the element is missing or not recognisable.

\subsubsection*{Exclusion considerations}
\begin{itemize}
    \item TCF incl: If the score is either 0 or 1 the data is included.
\end{itemize}

\subsubsection*{Outcome variables}
"TCF TotalScore", "TCF TotalTime".

\subsection*{Silhouetes}
Object recognition is measured with a short version of the silhouettes subtest of the VOSP \cite{vosp}, in which black-and-white silhouettes of five animals and five objects, presented from an unusual angle, have to be identified by the participant. 

\begin{figure}[H]
   \centering
   \includegraphics [width=0.5\textwidth] {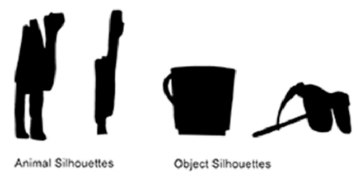}
  \caption{Examples of Silhouettes test presented to the subject.}
\end{figure}

\subsubsection*{Exclusion considerations}
\begin{itemize}
    \item SIL Completed: If the test was performed it is completed (e.g. 1) and the data is included.
\end{itemize}

\subsubsection*{Outcome variables}
"SIL CORR".

\subsection*{Crowding task}
The Crowding task (University of Groningen; department of developmental and clinical neuropsychology) is specifically designed for the DiaNAH test battery and consists of three conditions. The first two conditions are control conditions to assure the ability of the participant to identify isolated letters in the center and periphery and to deduce whether the participant fixates on the cross at the moment of stimulus presentation. Excessive crowding is assessed in the third and last condition. 

\begin{figure}[H]
   \centering
   \includegraphics [width=0.7\textwidth] {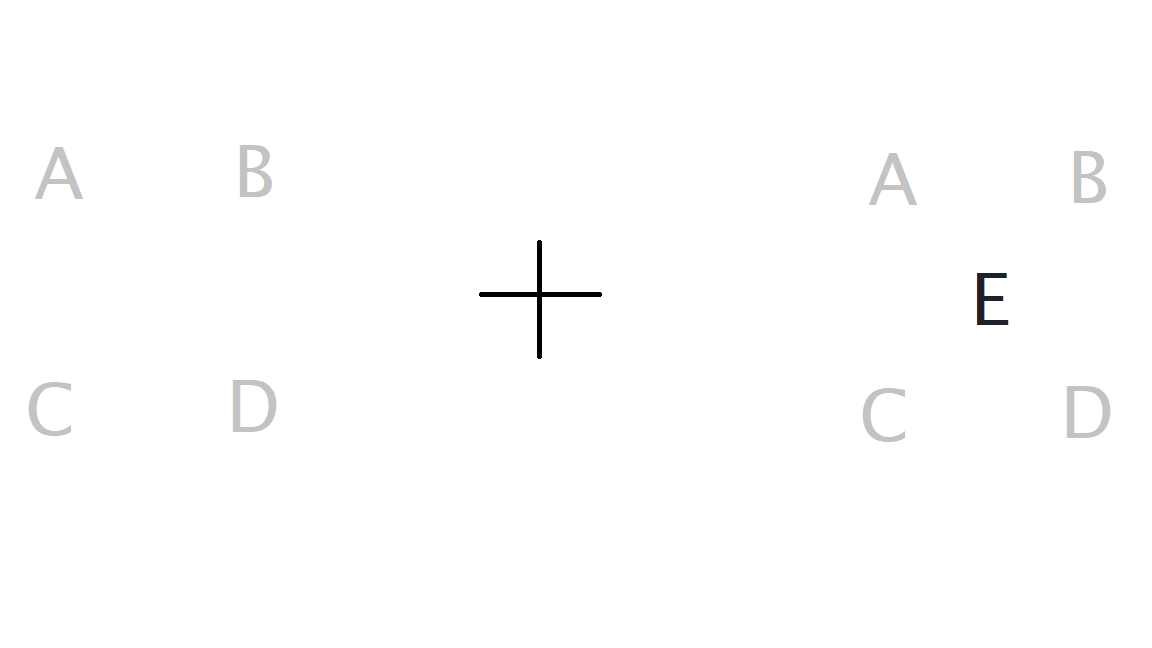}
  \caption{Reproduction of an example of the third condition of the Crowding task.}
\end{figure}

In all conditions, a fixation cross is presented in the center of the display. After 500 ms. the fixation cross disappears and the target stimulus appears, i.e. a letter (or letters, depending on the condition) in fixed width font (Lucida Sans, 80) and maximum contrast. The location of the target stimulus is condition dependent (i.e. central or in the periphery) but duration is the same for all conditions (150 ms).

If the participant is not able to perform the first two conditions, that is recognition of isolated letters in the center or the periphery is impaired, the third condition cannot be adequately interpreted.

\subsubsection*{Outcome variables}
"Crow cr Tcorr".

\section{Correlation Matrix}
\label{appendix:corrMatrix}
The correlation matrix $M$ can be computed for a dataset D with $n$ variables, where each entry $M[i][j]$ is the correlation coefficient between the $i$-th and $j$-th variable of D.

In mathematical notation, we denote the correlation coefficient between variables $X_i$ and $X_j$ as $corr(X_i, X_j)$. The correlation matrix $M$ can then be defined as:

\begin{equation}
    M = \begin{bmatrix}
    corr(X_1, X_1) & corr(X_1, X_2) & \dots & corr(X_1, X_n) \\
    corr(X_2, X_1) & corr(X_2, X_2) & \dots & corr(X_2, X_n) \\
    \vdots & \vdots & \ddots & \vdots \\
    corr(X_n, X_1) & corr(X_n, X_2) & \dots & corr(X_n, X_n) \\
    \end{bmatrix}
\label{eq:correlation_matrix}
\end{equation}

With
\begin{equation}
\text{corr}(X, Y) = \frac{\sum_{i=1}^{n} (x_i - \bar{x})(y_i - \bar{y})}{\sqrt{\sum_{i=1}^{n} (x_i - \bar{x})^2 \sum_{i=1}^{n} (y_i - \bar{y})^2}}
\end{equation}
, where $n$ is the number of pairs of scores, $x_i$ and $y_i$ are the individual sample points indexed with i, $\bar{x}$ and $\bar{y}$ are the means of x and y respectively.

The correlation matrix is a symmetric matrix as $corr(X_i, X_j) = corr(X_j, X_i)$, and the diagonal elements are always 1 as $corr(X_i, X_i) = 1$.

Regarding the interpretation of the correlation values:
\begin{itemize}
\item A correlation of 1 implies a perfect positive correlation.
\item A correlation of -1 indicates a perfect negative correlation.
\item A correlation close to 0 signifies no linear relationship between the variables.
\end{itemize}

\section{Power Analysis}
\label{appendix:poweranalysis}
\begin{align}
N_1 & = \frac{(z_{1-\frac{\alpha}{2}}\sqrt{\bar{p}\bar{q}(1+\frac{1}{\sqrt{K}})} + z_{1-\beta}\sqrt{p_1q_1+\frac{p_2q_2}{\sqrt{K}}})^2}{\Delta^2}
\label{eq:N_1}
\end{align}

Where, 
\begin{align}
\begin{split}
    q_1 & = 1-p_1 \\
q_2 & = 1-p_2 \\
\bar{p} & = \frac{p_1+kp_2}{1+K} \\
\bar{q} & = 1-\bar{p}
\end{split}
\label{eq:parameters}
\end{align}

With variables:
\begin{align*}
p_1, p_2 & : \text{proportion (incidence) of groups \#1 and \#2} \\
\Delta & : |p_2-p_1| = \text{absolute difference between two proportions} \\
N_1 & : \text{sample size for group \#1} \\
N_2 & : \text{sample size for group \#2} \\
\alpha & : \text{probability of type I error (usually 0.05)} \\
\beta & : \text{probability of type II error (usually 0.2)} \\
z & : \text{critical Z value for a given } \alpha \text{ or } \beta \\
K & : \text{ratio of sample size for group \#2 to group \#1}
\end{align*}

Solving equation \ref{eq:N_1} by plugging in the parameters listed in \ref{eq:parameters} yields:

\begin{align}
\begin{split}
N_1 & = \frac{(1.96\sqrt{0.5\cdot0.5\cdot(1+\frac{1}{\sqrt{1}})} + 0.84\sqrt{0.75\cdot0.25+\frac{0.25\cdot0.75}{\sqrt{1}}})^2}{0.5^2} \\
N_1 & = 14 \\
N_2 & = K\cdot N_1 = 14
\end{split}
\end{align}

\bibliography{main}

\begin{thebibliography}{12}
\expandafter\ifx\csname natexlab\endcsname\relax\def\natexlab#1{#1}\fi
\providecommand{\url}[1]{\texttt{#1}}
\providecommand{\href}[2]{#2}
\providecommand{\path}[1]{#1}
\providecommand{\DOIprefix}{doi:}
\providecommand{\ArXivprefix}{arXiv:}
\providecommand{\URLprefix}{URL: }
\providecommand{\Pubmedprefix}{pmid:}
\providecommand{\doi}[1]{\href{http://dx.doi.org/#1}{\path{#1}}}
\providecommand{\Pubmed}[1]{\href{pmid:#1}{\path{#1}}}
\providecommand{\bibinfo}[2]{#2}
\ifx\xfnm\relax \def\xfnm[#1]{\unskip,\space#1}\fi
\bibitem[{Carone(2007)}]{strauss}
\bibinfo{author}{Carone, D.} (\bibinfo{year}{2007}).
\newblock \bibinfo{title}{E. strauss, e. m. s. sherman, \& o. spreen, a
  compendium of neuropsychological tests: Administration, norms, and
  commentary}.
\newblock {\it \bibinfo{journal}{Applied Neuropsychology}\/},  {\it
  \bibinfo{volume}{14}\/}, \bibinfo{pages}{62--63}.
  \DOIprefix\doi{10.1080/09084280701280502}.
\bibitem[{Corsi(1973)}]{corsi}
\bibinfo{author}{Corsi, P.~M.} (\bibinfo{year}{1973}).
\newblock \bibinfo{title}{Human memory and the medial temporal region of the
  brain}.
\newblock {\it \bibinfo{journal}{Dissertation Abstracts International}\/},
  {\it \bibinfo{volume}{34}\/}, \bibinfo{pages}{891}.
\bibitem[{Gauthier et~al.(1989)Gauthier, Dehaut \& Joanette}]{Gauthier1989}
\bibinfo{author}{Gauthier, L.}, \bibinfo{author}{Dehaut, F.}, \&
  \bibinfo{author}{Joanette, Y.} (\bibinfo{year}{1989}).
\newblock \bibinfo{title}{Bells test}.
\newblock \URLprefix \url{https://doi.org/10.1037/t28075-000}.
  \DOIprefix\doi{10.1037/t28075-000}.
\bibitem[{Howieson \& Lezak(1995)}]{lezak}
\bibinfo{author}{Howieson, D.~B.}, \& \bibinfo{author}{Lezak, M.~D.}
  (\bibinfo{year}{1995}).
\newblock \bibinfo{title}{Separating memory from other cognitive problems}.
\newblock In \bibinfo{editor}{A.~D. Baddeley}, \bibinfo{editor}{B.~A. Wilson},
  \& \bibinfo{editor}{F.~N. Watts} (Eds.), {\it \bibinfo{booktitle}{Handbook of
  Memory Disorders}\/} (pp. \bibinfo{pages}{411--426}).
\newblock \bibinfo{publisher}{John Wiley \& Sons}.
\bibitem[{MacPherson et~al.(2019)MacPherson, Allerhand, Cox \&
  Deary}]{macpherson}
\bibinfo{author}{MacPherson, S.~E.}, \bibinfo{author}{Allerhand, M.},
  \bibinfo{author}{Cox, S.~R.}, \& \bibinfo{author}{Deary, I.~J.}
  (\bibinfo{year}{2019}).
\newblock \bibinfo{title}{Individual differences in cognitive processes
  underlying trail making test-b performance in old age: The lothian birth
  cohort 1936}.
\newblock {\it \bibinfo{journal}{Intelligence}\/},  {\it
  \bibinfo{volume}{75}\/}, \bibinfo{pages}{23--32}.
\bibitem[{Milner(1971)}]{milner}
\bibinfo{author}{Milner, B.} (\bibinfo{year}{1971}).
\newblock \bibinfo{title}{Interhemispheric differences in the localization of
  psychological processes in man.}
\newblock {\it \bibinfo{journal}{British medical bulletin}\/}, .
\bibitem[{Posner et~al.(1969)Posner, Boies, Eichelman \& Taylor}]{taylor}
\bibinfo{author}{Posner, M.~I.}, \bibinfo{author}{Boies, S.~J.},
  \bibinfo{author}{Eichelman, W.~H.}, \& \bibinfo{author}{Taylor, R.~L.}
  (\bibinfo{year}{1969}).
\newblock \bibinfo{title}{Retention of visual and name codes of single
  letters.}
\newblock {\it \bibinfo{journal}{Journal of experimental psychology}\/},  {\it
  \bibinfo{volume}{79}\/}, \bibinfo{pages}{1}.
\bibitem[{Reitan \& Wolfson(2004)}]{reiten}
\bibinfo{author}{Reitan, R.~M.}, \& \bibinfo{author}{Wolfson, D.}
  (\bibinfo{year}{2004}).
\newblock \bibinfo{title}{The trail making test as an initial screening
  procedure for neuropsychological impairment in older children}.
\newblock {\it \bibinfo{journal}{Archives of Clinical Neuropsychology}\/},
  {\it \bibinfo{volume}{19}\/}, \bibinfo{pages}{281--288}.
\bibitem[{Torfs et~al.(2013)Torfs, Vancleef, Lafosse, Wagemans \&
  de~Wit}]{Torfs2013TheLP}
\bibinfo{author}{Torfs, K.}, \bibinfo{author}{Vancleef, K.},
  \bibinfo{author}{Lafosse, C.}, \bibinfo{author}{Wagemans, J.}, \&
  \bibinfo{author}{de~Wit, L.} (\bibinfo{year}{2013}).
\newblock \bibinfo{title}{The leuven perceptual organization screening test
  (l-post), an online test to assess mid-level visual perception}.
\newblock {\it \bibinfo{journal}{Behavior Research Methods}\/},  {\it
  \bibinfo{volume}{46}\/}, \bibinfo{pages}{472 -- 487}. \URLprefix
  \url{https://api.semanticscholar.org/CorpusID:18320830}.
\bibitem[{Vancleef et~al.(2015)Vancleef, Acke, Torfs, Demeyere, Lafosse,
  Humphreys, Wagemans \& de~Wit}]{vancleef}
\bibinfo{author}{Vancleef, K.}, \bibinfo{author}{Acke, E.},
  \bibinfo{author}{Torfs, K.}, \bibinfo{author}{Demeyere, N.},
  \bibinfo{author}{Lafosse, C.}, \bibinfo{author}{Humphreys, G.},
  \bibinfo{author}{Wagemans, J.}, \& \bibinfo{author}{de~Wit, L.}
  (\bibinfo{year}{2015}).
\newblock \bibinfo{title}{Reliability and validity of the l euven perceptual
  organization screening test (l-post)}.
\newblock {\it \bibinfo{journal}{Journal of Neuropsychology}\/},  {\it
  \bibinfo{volume}{9}\/}, \bibinfo{pages}{271--298}.
\bibitem[{de~Vries et~al.(2022)de~Vries, Tucha, Melis-Dankers, Vrijling,
  Ribbers, Cornelissen \& Heutink}]{de2022birthday}
\bibinfo{author}{de~Vries, S.}, \bibinfo{author}{Tucha, O.},
  \bibinfo{author}{Melis-Dankers, B.}, \bibinfo{author}{Vrijling, A.},
  \bibinfo{author}{Ribbers, S.}, \bibinfo{author}{Cornelissen, F.}, \&
  \bibinfo{author}{Heutink, J.} (\bibinfo{year}{2022}).
\newblock \bibinfo{title}{The birthday party test (bpt): A new picture
  description test to support the assessment of simultanagnosia in patients
  with acquired brain injury}.
\newblock {\it \bibinfo{journal}{Applied Neuropsychology: Adult}\/},  {\it
  \bibinfo{volume}{29}\/}, \bibinfo{pages}{383--396}.
\bibitem[{Warrington \& James(1991)}]{vosp}
\bibinfo{author}{Warrington, E.}, \& \bibinfo{author}{James, M.}
  (\bibinfo{year}{1991}).
\newblock {\it \bibinfo{title}{The Visual Object and Space Perception Battery:
  VOSP}\/}.
\newblock \bibinfo{publisher}{Pearson}.
\newblock \URLprefix \url{https://books.google.nl/books?id=znV_PQAACAAJ}.

\end{thebibliography}

\end{document}